# SIMULATION OF BORON DIFFUSION DURING ANNEALING OF SILICON SUBSTRATES UNDERGONE A HIGH FLUENCE ION IMPLANTATION

## O. I. Velichko and A. A. Hundoryna

**Abstract**

A theoretical investigation of the microscopic mechanisms provided the transient enhanced diffusion of boron atoms during rapid thermal annealing of silicon substrates doped by high fluence ion implantation was carried out. To compare the mechanisms a model of the transient enhanced diffusion due to migration of the pairs "boron atom — silicon interstitial" was developed. It is supposed that during annealing dissolution of the clusters incorporated boron atoms occurs. During cluster dissolution, a fraction of boron atoms occupies a substitutional position, whereas other atoms become interstitial. It was shown from the comparison of the shape of calculated boron concentration profile after annealing with the experimental data that at a temperature of 850 Celcius degrees and below the nonequilibrium boron interstitials are responsible for the transient enhanced diffusion. On the other hand, at a temperature of 850 Celcius degrees and above a major contribution to the transient enhanced diffusion is provided by the pairs "boron atom — silicon interstitial".

## ВВЕДЕНИЕ

В настоящее время для производства кремниевых интегральных микросхем (ИМС) широко используется низкоэнергетическая высокодозная ионная имплантация в сочетании с короткими термическими отжигами для удаления постимплантационных дефектов. В процессе термообработки ионно-имплантированных слоев имеет место скоротечная диффузия, которая является в настоящее время основным лимитирующим фактором при повышении степени интеграции ИМС. Особенно сильно скоротечная диффузия проявляется случае перераспределения бора, имплантированного в кремний с дозами, обеспечивающими создание высоколегированного слоя с концентрацией примеси выше предела растворимости при используемой температуре обработки.

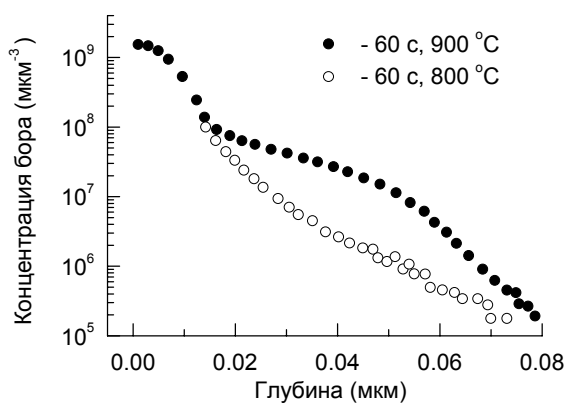

Рисунок 1 — измеренные распределения концентрации бора в ионно-имплантированных подложках, подвергнутых термическому отжигу в течении 60 с при 800 и 900 °C. Экспериментальные данные взяты из работы [1].

На Рисунке 1 представлены профили распределения ионно-имплантированного бора после термообработок одинаковой длительности при температурах 800 и 900 °C, измеренные в

[1] методом вторичной ионной масс-спектроскопии (ВИМС). В работе [1] имплантация ионов бора с энергией 1 кэВ и дозой $1.5 \times 10^{15}$ см$^{-2}$ осуществлялась в кремниевые подложки $n$-типа проводимости ориентации (100). Приповерхностный слой кремния предварительно аморфизовывался имплантацией ионов Ge с энергией 1 кэВ и дозой $1.5 \times 10^{15}$ см$^{-2}$. После имплантации бора осуществлялся быстрый термический отжиг в атмосфере азота длительностью 60 секунд при различных температурах. от 650 °C до 1000 °C.

Как видно из Рисунка 1, профили ионно-имплантированной примеси претерпевают существенное диффузионное перераспределение, которое выражается в переносе значительной части примесных атомов вглубь полупроводника. Используя данные Рисунка 1, можно сформулировать следующие особенности профилей распределения ионно-имплантированного бора после быстрого термического отжига:

1. Основная часть атомов бора в области высокой концентрации примесных атомов остается неподвижной. В области максимальной концентрации примесных атомов имеет место лишь незначительное перераспределение бора, что приводит к сохранению пика концентрации примеси, сформированного ионной имплантацией.

2. Наблюдается существенное перераспределение бора в области концентраций меньших $10^8$ мкм$^{-3}$.

3. Перераспределение бора в области низких концентраций носит различный характер для рассматриваемых температур 800 и 900 °C:

а) Так, для температуры 800 °C концентрационный профиль распределения бора имеет вид протяженного "хвоста", направленного вглубь полупроводника, причем этот "хвост" близок к прямой линии в логарифмическом масштабе по оси концентраций.

б) Как видно из Рисунка 1, после отжига при температуре 900 °C профиль распределения бора в области концентраций меньших $10^8$ мкм$^{-3}$ становиться выпуклым, причем большая доля примеси находится в области концентраций от $10^8$ до $10^7$ мкм$^{-3}$.

Можно предположить, что различный вид профилей распределения атомов бора для температур 800 и 900 °C связан с изменением микроскопического механизма диффузии при повышении температуры.

**Цель исследования**

Предположение, описанное выше, позволяет сформулировать цель настоящего исследования — определить микроскопические механизмы переноса атомов бора в диапазоне температур 800... — 900 °C, когда имеет место переход от одного механизма диффузии к другому.

**Анализ микроскопических механизмов переноса атомов примеси**

В работе [2] было высказано предположение, что образование протяженных "хвостов" в области низкой концентрации примесных атомов, которое имеет место при перераспределении ионно-имплантированной примеси в случае коротких низкотемпературных термообработок, есть следствие длиннопробежной миграции неравновесных межузельных атомов примеси. Аналитическое решение краевой задачи диффузии примесных атомов для случая непрерывной генерации неравновесных межузельных атомов примеси в пределах имплантированного слоя было получено в работе [3]. Предполагалось, что может иметь место испарение межузельных атомов примеси с поверхности полупроводника. Поэтому, на поверхности задавалось наиболее общее граничное условие 3-го рода, тогда как в объеме полупроводника задавалось традиционное условие постоянства (равенства нулю) концентрации межузельных атомов примеси. Анализ полученного аналитического решения показывает, что в процессе отжига происходит формирование "хвоста" в области низкой концентрации примесных атомов, причем концентрационный профиль распределения атомов примеси в этой области представляет собой прямую линию при логарифмическом масштабе по оси концентраций. Этот вывод подтверждается результатами расчетов работы [4].

Таким образом, можно предположить, что в рассматриваемом случае при температурах около и ниже 800 $^{o}$C атомы ионно-имплантированного бора, которые переходят в положение замещения в результате твердофазной рекристаллизации аморфного слоя, остаются неподвижными в течение всего времени отжига. Причиной скоротечной диффузии бора, которая приводит к формированию протяженного "хвоста", направленного вглубь полупроводника, при данных температурах обработки является длиннопробежная миграции неравновесных межузельных атомов примеси. Как было предположено в работе [4], такие неравновесные межузельные атомы бора могут генерироваться в ионно-имплантированном слое вследствие образования (перестройки) кластеров атомов бора с собственными точечными дефектами или в результате деформации кристалла, обусловленной разницей атомных радиусов бора и кремния.

Для объяснения выпуклой формы концентрационного профиля распределения атомов бора естественно предположить, что при температурах 900 $^{o}$C и выше начинают диффундировать атомы примеси, находящиеся в положении замещения. Считается, что в этом случае диффузия бора осуществляется посредством образования, миграции и распада пар "примесный атом — межузельный атом кремния" [5,6]. В ионно-имплантированных слоях диффузия бора посредством пар усиливается наличием большого количества неравновесных межузельных атомов кремния. Эти межузельные атомы могут генерироваться вследствие отжига постимплантационных дефектов за пределами рекристаллизованного слоя, а также в результате распада кластеров атомов бора, включающих межузельные атомы кремния [7].

**Модель диффузии**

Как видно из Рисунка 1, при температуре 900 $^{o}$C основная часть атомов бора в области высокой концентрации примеси остается неподвижной, образуя кластеры атомов бора с межузельными атомами кремния [7]. При термообработке происходит распад этих кластеров, в результате чего атомы бора переходят в положение замещения и могут диффундировать по механизму образования, миграции и распада пар. С учетом распада кластеров атомов примеси для описания диффузии атомов бора, находящихся в положении замещения, можно использовать следующее уравнение [8]:

$$\frac{\partial C^T}{\partial t} = \nabla \left\{ D^I(\chi) \left[ \nabla(\tilde{C}^I C) + \tilde{C}^I C \frac{\nabla \chi}{\chi} \right] \right\} + G(x) , \qquad (1)$$

где

$$\tilde{C}^I = \frac{C^I}{C_i^I} .$$

Здесь $C^T$ и $C = C(x,t)$ — общая концентрация бора и концентрация атомов бора в положении замещения, соответственно; $C^I$ и $C_i^I$ — концентрация межузельных атомов кремния и термически равновесное значение этой концентрации, соответственно; $D^I(\chi)$ — коэффициент диффузии бора; $\chi$ — концентрация дырок, приведенная к концентрации собственных носителей заряда в полупроводнике $n_i$ для данной температуры обработки; $G(x)$ — скорость перехода атомов примеси в положение замещения при распаде кластеров атомов бора.

Для численного решения уравнения диффузии (1) зададим начальное условие

$$C(x,0) = C_0(x),$$

где $C_0(x)$ — распределение атомов бора в положении замещения до начала отжига, а также два граничных условия, а именно: граничное условие третьего рода на поверхности и граничное условие первого рода в объеме полупроводника.

### Результаты численных расчетов

Проведем численные расчеты для процесса диффузии при температуре 850 °C, исследованного в [9], поскольку в этой работе измерялась концентрация атомов электрически активного бора. На Рисунке 2 представлены профили распределения общей концентрации примеси, измеренной методом ВИМС, и концентрации электрически активного бора, определенной методом измерения сопротивления растекания в сочетании с послойным стравливанием.

В работе [9] имплантация ионов бора осуществлялась с энергией 1.5 кэВ и дозой $3 \times 10^{15}$ см$^{-2}$ в кремниевые подложки $n$-типа проводимости с ориентацией (100). Преаморфизация кремния осуществлялась посредством имплантацией ионов германия с энергией 12 кэВ и дозой $1 \times 10^{15}$ см$^{-2}$. В результате имплантации германия создавался аморфный слой толщиной 22 нм, то есть большая часть имплантированного бора приходилась на аморфную область. После имплантации бора осуществлялся быстрый термический отжиг длительностью 60 секунд. Профиль распределения электрически активного бора после отжига при 850 °C представлен на Рисунке 2 зачерненными кружками.

При расчете были использованы следующие значения параметров имплантации и диффузии атомов бора: $R_p$ = 0.00674 мкм, $\Delta R_p$ = 0.00476 мкм; $Sk$ = 0.275; $D^I(\chi)$ = $3.4 \times 10^{-8}$ мкм$^2$/с. Здесь $R_p$ и $\Delta R_p$ — средний проективный пробег ионов и страгглинг этого пробега соответственно; $Sk$ — асимметрия профиля распределения имплантированной примеси.

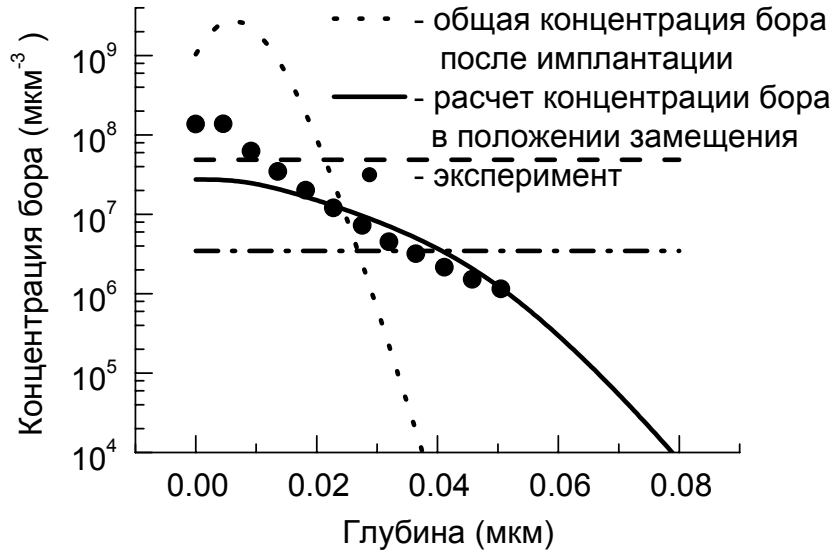

Рисунок 2 — рассчитанное распределение концентрации атомов бора в положении замещения. Экспериментальные данные взяты из работы [9].
Штриховая кривая — предел растворимости бора в кремнии
Штрих-пунктирная кривая — концентрация собственных носителей заряда $n_i$

Как видно из Рисунка 2, рассчитанный профиль распределения электрически активного бора после отжига имеет выпуклую форму. Это означает, что предложенная модель может объяснить особенности профиля распределения бора, которые наблюдаются в случае отжига при 900 $^o$С. В тоже время экспериментальные данные показывают, что профиль распределения бора в области "хвоста", направленного вглубь полупроводника, близок к прямой линии в логарифмическом масштабе по оси концентраций. Это означает, что при температуре отжига 850 $^o$С основную роль в скоротечной диффузии играет длиннопробежная миграции неравновесных межузельных атомов примеси, которые генерируются в области высокой концентрации имплантированной примеси в результате распада или перестройки кластеров атомов бора.

**СПИСОК ЛИТЕРАТУРЫ**